\documentclass[prl, aps, showpacs,superscriptaddress,twocolumn,longbibliography]{revtex4-2}

\usepackage{hyperref}
\hypersetup{
     colorlinks=true,
     linkcolor=magenta,
     filecolor=blue,
     citecolor=blue,      
     urlcolor=cyan,
     }
\usepackage{color}
\usepackage[usenames,dvipsnames]{xcolor}
\usepackage{amsmath,amsthm,amssymb}
\usepackage{graphicx}
\usepackage{float}
\usepackage{epsfig}
\usepackage{bm}% bold math
\usepackage{mathrsfs}
\usepackage{multirow}
\usepackage[all]{xy}
\usepackage{pbox}
\usepackage{verbatim}
\usepackage{braket}
\usepackage{mathtools}
\usepackage{bm}
\usepackage{tikz}
\usepackage{xcolor}
 
\usepackage{mathtools}

\usepackage{mathtools}
\usepackage{tabstackengine}
\usepackage{enumerate}   
\usepackage[normalem]{ulem}
\stackMath

\newcommand{\er}[1]{Eq.~\eqref{#1}}
\newcommand{\ers}[2]{Eqs.~(\ref{#1}-\ref{#2})}

\newcommand{\beq}{\begin{equation}}
\newcommand{\eeq}{\end{equation}}

\begin{document}  

\title{Optimal sampling of dynamical large deviations in two dimensions via tensor networks}

\author{Luke Causer}
\affiliation{School of Physics and Astronomy, University of Nottingham, Nottingham, NG7 2RD, UK}
\affiliation{Centre for the Mathematics and Theoretical Physics of Quantum Non-Equilibrium Systems,
University of Nottingham, Nottingham, NG7 2RD, UK}
\author{Mari Carmen Ba\~nuls}
\affiliation{Max-Planck-Institut f\"ur Quantenoptik, Hans-Kopfermann-Str.\ 1, D-85748 Garching, Germany}
\affiliation{Munich Center for Quantum Science and Technology (MCQST), Schellingstr.\ 4, D-80799 M\"unchen}
\author{Juan P. Garrahan}
\affiliation{School of Physics and Astronomy, University of Nottingham, Nottingham, NG7 2RD, UK}
\affiliation{Centre for the Mathematics and Theoretical Physics of Quantum Non-Equilibrium Systems,
University of Nottingham, Nottingham, NG7 2RD, UK}

\begin{abstract}
We use projected entangled-pair states (PEPS) to calculate the large deviation statistics of the dynamical activity of the two dimensional East model, and the two dimensional symmetric simple exclusion process (SSEP) with open boundaries, in lattices of up to $40\times 40$ sites. 
We show that at long times both models have phase transitions between active and inactive dynamical phases.
For the 2D East model we find that this trajectory transition is of the first-order, while for the SSEP we find indications of a second order transition. We then show how the PEPS can be used to implement a trajectory sampling scheme capable of directly accessing rare trajectories.  We also discuss how the methods described here can be extended to study rare events at finite times.
\end{abstract}

\maketitle

\textbf{\em Introduction.-}
Tensor network (TN) techniques, whereas most actively developed in the context of quantum many-body physics~\cite{Verstraete2008,Schollwoeck2011,Orus2014annphys,Silvi2019tn,Okunishi2022,Banuls2023}, 
offer powerful numerical tools for much more general problems.
They are based on an efficient parametrization of the many-body state in terms of local tensors (multidimensional arrays) connected according to a graph that, in general, responds to 
the structure of correlations in the state. 
In the last few years we have seen progress in their application
to compute statistical properties of dynamical trajectories in classical stochastic systems. The first application was to the long time statistics---the dynamical large deviation (LD) regime---of one-dimensional lattice systems 
using variational algorithms (such as density matrix renormalization group \cite{White1992}, or DMRG) to approximate the leading eigenvectors of tilted Markov generators by matrix product states (or MPS, see e.g.\ Ref.~\cite{Schollwoeck2011})
\cite{Gorissen2009, Gorissen2012, Gorissen2012a, Banuls2019, Helms2019, Causer2020, Causer2022a, Gu2022}. Building on these results, we recently introduced (i) a method which exploited MPS to efficiently sample long-time rare trajectories, and (ii) an MPS time-evolution to precisely compute trajectory statistics at finite times \cite{Causer2022}.

The suitability of the TN ansatz for these problems is rooted in the fact that the targeted eigenvectors have low correlations. 
More concretely, for local problems, we expect them to fulfill (up to small corrections) a so-called area law~\cite{Eisert2010},
according to which the ``entanglement'' (or a mathematically analogous quantity for classical probability vectors~\cite{Prosen2007osee})
with respect to a bipartition is upper-bounded by the size of its boundary.
This scaling is captured by the MPS family in one spatial dimension. 
In higher dimensions, a suitable generalization with area law is provided by
%For higher dimensions one needs a generalisation, such as  
the projected-entangled pair states (PEPS)~\cite{Verstraete2004},
which 
were recently applied to the classical asymmetric exclusion process in two dimensions in Ref.~\cite{Helms2020}. 
A computationally cheaper alternative, without an area law but accommodating more entanglement than MPS, is that of tree tensor networks \cite{Shi2006}, used for example in Refs.~\cite{Strand2022, Strand2022a} (in combination with a time-dependent variational principle~\cite{Bauernfeind2020}) to study driven problems.

Here we use PEPS to study the LDs of the {\em dynamical activity} in two paradigmatic two-dimensional models, 
the 2D East model (also known as North-or-East model) \cite{Jaeckle1991,Ritort2003,Berthier2005,Casert2021}, and the 2D symmetric simple exclusion process (SSEP) with open boundaries where particles can be injected and removed \cite{Mallick2015}.
We are able to accurately estimate the leading eigenvector of the tilted generator - and thus the LDs - of these models, and construct a close-to-optimal dynamics to directly sample the corresponding rare trajectories. Such an algorithm requires efficient sampling from the PEPS, and we show how to do this in the context of trajectory sampling. We benchmark our methods, showing how the PEPS allows for a controlled accuracy of optimal dynamics. We demonstrate that both models have a phase transition between active and inactive dynamical phases, a first-order transition for the 2D East and a second-order transition for the 2D SSEP. 

\begin{figure}[ht]
    \centering
    \includegraphics[width=\linewidth]{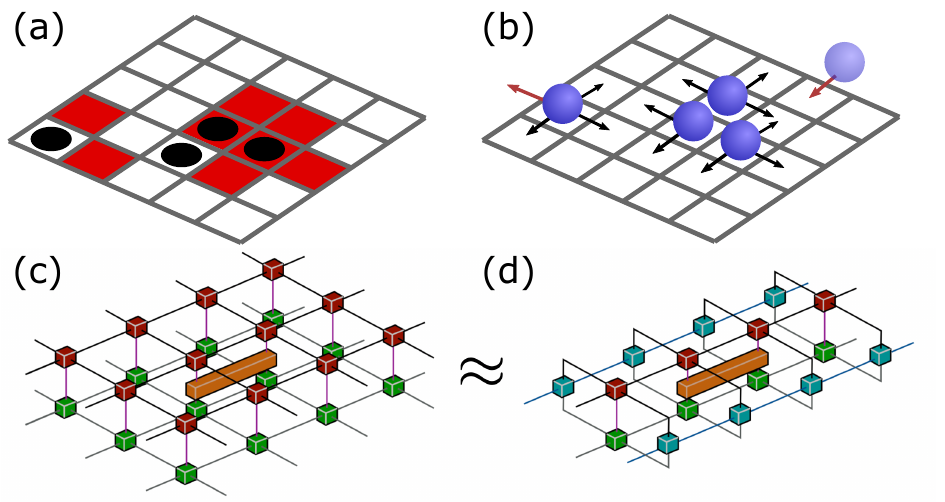}
    \caption{
        {\bf Models.}
        (a) 2D East: an occupied site (black circles) facilitates flips in neighboring sites (red cells) only in two directions. 
        (b) 2D SSEP: particles can hop to empty neighboring sites (black arrows), symmetrically in any direction; particles can enter or leave at the boundaries (red arrows).
        {\bf PEPS.}
        (c) A PEPS is parametrized by a tensor per lattice site (red boxes for the top PEPS), each local tensor with a physical index (purple line) and four virtual legs (black lines) that connect it to  neighboring tensors. The expectation value of a local operator (orange box) is obtained by sandwiching it between the PEPS and its adjoint (shown in green), and contracting (i.e. multiplying and summing over) the physical (basis) indices. (d) The cost of exact contraction scales exponentially with size, so an MPS approximation (blue tensors) is used for the contraction part of the network, with the dimension of its virtual bonds (blue legs) controlling accuracy; see Ref.~\cite{SM} for details.
    }
    \label{fig: models}
\end{figure}

\textbf{\em Models.-}
The models we study here live in a  two-dimensional square lattice of size $N = L \times L$, with each site being occupied by a binary variable $n_{\bm k} = 0$ or $1$, where ${\bm k} = (k_{x}, k_{y})$ denotes the position of the site for $k_{x}, k_{y} = 1\cdots L$. Their continuous-time dynamics is defined by a Markov generator (e.g. see Refs.~\cite{Gardiner2004,Garrahan2018}),
\beq
    \mathbb{W} = \sum_{x, y\neq x} w_{x\to y}\ket{y}\bra{x} - \sum_{x} R_{x} \ket{x}\bra{x},
\eeq
where $\ket{x}$ and $\ket{y}$ are configurations on the lattice, $w_{x\to y}$ the transition rate from $x$ to $y$, and
$R_{x} = \sum_{y\neq x} w_{x\to y}$ the escape rate out of $x$.
We can write this as $\mathbb{W} = \mathbb{K} - \mathbb{R}$, where $\mathbb{K}$ contains the off-diagonal transition rates, and $\mathbb{R}$ the diagonal escape rates.

The first model we consider is the 2D East model \cite{Jaeckle1991,Ritort2003,Berthier2005,Casert2021}, often studied in the context of the glass transition. 
This is a kinetically constrained model such that an excited site $n_{\bm k} = 1$, allows (``facilitates'') a site to its North or East to flip stochastically, see Fig.~\ref{fig: models}(a).
It is parametrized by $c \in (0, 1/2]$, which determines the local transitions rates: $0\to 1$ with rate $c$, and $1 \to 0$ with rate $1-c$, subject to the kinetic constraint.
In addition, we choose open boundary conditions with $n_{(1, 1)} = 1$ fixed.
This ensures the entire state space remains dynamically connected \cite{Ritort2003}.
The second model is the 2D SSEP.
This describes particles hopping to neighbouring sites on a 2D lattice with unit rate,
but only if the target site is not already occupied by a particle.
We also allow particles to be injected or removed at the boundaries of the lattice with rate $1/2$, see Fig.~\ref{fig: models}(b).
Exact definitions of the models are given in Ref.~\cite{SM}.

\textbf{\em Dynamical large deviations.-}
We consider the statistics of a dynamical observable $\hat{K}$ through its probability distribution
$
    P_{t}(K) = \sum_{\omega_{t}} \pi(\omega_{t}) \delta[\hat{K}(\omega_{t}) - K],
$
where $\omega_{t}$ denotes a stochastic trajectory and $\pi(\omega_{t})$ its probability. The corresponding moment generating function is
$
    Z_{t}(s) = \sum_{\omega_{t}} \pi(\omega_{t})e^{-s\hat{K}(\omega_{t})} .
$
In the $t\to\infty$ limit, the two obey LD principles 
$P_{t}(K) \asymp e^{-t\varphi(K/t)}$
and
$Z_{t}(s) \asymp e^{t\theta(s)}$,
with the rate function $\varphi(K/t)$ and scaled cumulant generating function (SCGF) $\theta(s)$ related through a Legendre transform,
$
    \theta(s) = -\min_{k}\left[sk + \varphi(k)\right]
$
for $k = K / t$ 
(for reviews see Refs.~\cite{Touchette2009, Touchette2018, Garrahan2018, Jack2020}.)

We focus as an observable on the {\em dynamical activity} \cite{Garrahan2007,Maes2020}, which counts the number of jumps in a trajectory. The relevant operator to study is the {\em tilted generator} \cite{Touchette2009, Touchette2018, Garrahan2018, Jack2020}, which for the activity reads
$
    \mathbb{W}_{s} = e^{-s}\mathbb{K} - \mathbb{R}
$, with the LD statistics encoded in the leading eigenvalue and (right and left) eigenvector(s), $\mathbb{W}_{s}\ket{r_{s}} = \theta(s)\ket{r_{s}}$ 
and    
$\bra{l_{s}}\mathbb{W}_{s} = \theta(s)\bra{l_{s}}$.

% \begin{figure}[t]
%     \centering
%     \includegraphics[width=\linewidth]{su_vu.pdf}
%     \caption{\textbf{Optimization of PEPS.} 
%     The error in measured energy for the SU and the FU compared to the high accuracy 2D DMRG (with a MPS bond dimension up to $D_{\rm MPS} = 1024$) for various values of $s$ and a $10\times10$ lattice. 
%     The left panel shows the 2D East model with $c = 0.5$, and the right panel shows the 2D SSEP.
%     The PEPS environment in the FU uses a boundary dimension $\chi_{B} = 4D^2$ for the East and $\chi_{B} = 6D^2$ for the SSEP.
%     }
%     \label{fig: peps_error}
% \end{figure}

\begin{figure*}[t]
    \centering
    \includegraphics[width=\linewidth]{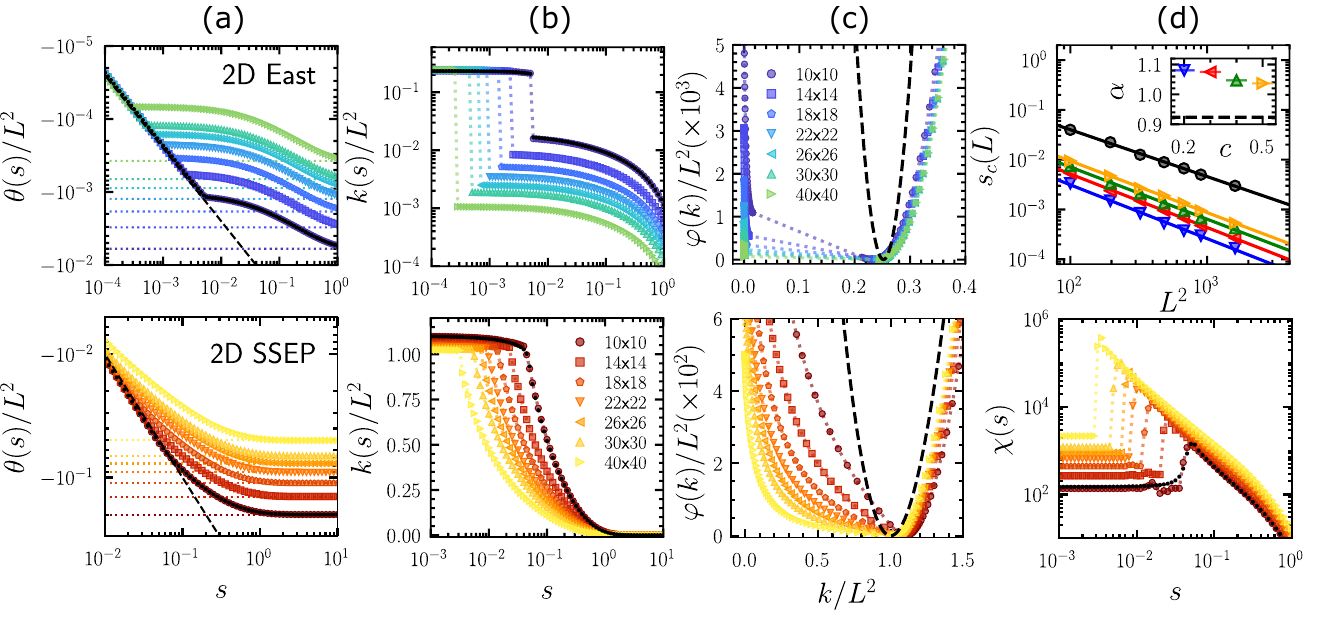}
    \caption{\textbf{Dynamical large deviations and active-inactive transitions from PEPS.} 
    (a) The SCGF $\theta(s)/L^{2}$ for the 2D East  with $c = 0.3$ (top) and the SSEP (bottom) for system sizes $N\in[10^2, 40^2]$. The black dashed line shows the linear response for small $s$, and the colour dotted lines show the value for $s\to\infty$.
    (b) The dynamical activity $k(s)/L^{2}$ for the systems in (a). The East is on a log-log scale, and the SSEP a log-linear scale.
    (c) The rate function $\varphi(k)/L^{2}$ as a function of activity $k / L^{2}$ for the systems in (a). The dashed line shows the Poisson distribution with mean $k(s = 0) / L^{2}$.
    (d) The transition points $s_{c}(L)$ for the 2D SSEP (black circles) and the 2D East for $c \in [0.2, 0.5]$. 
    The solid lines show the fitted power-law curves $s_{c}(L) \sim L^{-2\alpha}$, with the exponents shown in the inset. The black dashed line is the exponent for the SSEP, and the symbols are for the East. The symbols can be used to read the value of $c$ in the main figure.
    The bottom panel shows the dynamical susceptibility $\chi(s) = \theta''(s)$ for the 2D SSEP.
    All the data was acquired using the SU except for the black markers, which show (quasi-exact) 2D DMRG data for a $N = 10$ lattice for comparison.
    }
    \label{fig: large_deviations}
\end{figure*}

\textbf{\em Projected-entangled pair states.-}
TN methods allow us to solve the problem above using a variational ansatz for $\ket{\psi_{s}}$ in the PEPS family, a
natural generalization of MPS for area law states and lattices in more than one spatial dimension~\cite{Verstraete2004}. 
PEPS parametrize the many-body state with one rank-5 tensor per lattice site, in which the \emph{physical} index has the dimension of the site variable ($0,1$ in our case), and the remaining four \emph{virtual} indices each have bond dimension $D_{\rm PEPS}$, determining the maximum number of parameters in the ansatz, $N_{p} = NdD_{\rm PEPS}^{4}$
(see Fig.~\ref{fig: models} for a graphical representation).
Several TN algorithms exist to optimize the PEPS by maximizing the expectation of a local stochastic generator.
Crucial to all of them is an efficient computation of expectation values for local operators, such as the terms in $\mathbb{W}_{s}$.
We use the boundary MPS scheme~\cite{Verstraete2004, Verstraete2008} (illustrated in Fig.~\ref{fig: models}),
which approximates the partial contraction of the network by a MPS, whose bond dimension $\chi_{B}$ controls the accuracy of the contraction.
A heuristic choice for local problems is $\chi_{B} \sim O(D_{\rm PEPS}^{2})$ (see e.g. Ref.~\cite{Lubasch2014unifying}).

In order to find the PEPS approximation to the leading eigenvector $\ket{\psi_{s}}$ we employ time evolution,
which effectively projects the ansatz onto the leading eigenvector by iteratively applying short evolution steps, 
decomposed in two-body terms that are applied sequentially~\cite{Lubasch2014,Banuls2023}.
After each operation, the directly affected pair of tensors is updated. 
This requires a strategy to approximate their environment (i.e. the contraction of the remaining TN).
After comparing to quasi-exact
\footnote{The results are compared to those of 2D DMRG. They are considered quasi-exact, because by checking for covergence in the MPS bond dimension, we ensure high precision in the measured value, with an error which is orders of magnitude smaller than those of PEPS.}
results and to a more expensive strategy, 
we find that the computationally cheapest simple update (SU)~\cite{Jiang2008}, with maximal PEPS bond dimension $D_{\rm PEPS} = 4$, is enough to achieve
sufficiently accurate measurements of the SCGF in our problems, and allows us reaching large sizes at low computational cost, scaling only as $O(D_{\rm PEPS}^5)$.
For further details on the numerical approach we use see Ref.~\cite{SM}.

\textbf{\em Large deviations from PEPS.-}
The East and SSEP in 1D are known to have dynamical phase transitions in terms of the activity or other dynamical observables 
\cite{Garrahan2007, Garrahan2009, Derrida2007, Bodineau2007,AppertRolland2008,Lecomte2012,Jack2015,Nemoto2017, Karevski2017,Banuls2019, Causer2020}. In two-dimensions, 
the SSEP has a transition in the LDs of the current \cite{Helms2020}.
We now provide evidence by means of PEPS for both the 2D East and 2D SSEP having active-inactive phase transitions. Figure \ref{fig: large_deviations}(a-c) shows the LD statistics for both the 2D East model (top) and the 2D SSEP (bottom).
For the East model, we see from Fig.~\ref{fig: large_deviations}(a) that the SCGF follows a linear response, $\theta(s) \approx sk(0)$, for small $s$, but at $s_{c}(L)$ it sharply changes to another branch. 
This point corresponds to a sudden drop in activity, $k(s) = -\theta'(s)$, which becomes discontinuous in the limit $N\to\infty$, see Fig.~\ref{fig: large_deviations}(b). 
Having access to both the SCGF and the dynamical activity allows us to estimate the rate function $\varphi(k)$, shown in Fig.~\ref{fig: large_deviations}(c).
We see a broadening of the rate function around the mean, indicating the coexistence of active and inactive dynamics. 
For comparison, we also show the distribution of a simple process with the same mean activity, but which is uncorrelated in time (black dashed line).
All this behaviour is characteristic of a first-order phase transition.

For the SSEP we see something different:  
Fig.~\ref{fig: large_deviations}(a) shows no sharp change in $\theta(s)$, and the activity in Fig.~\ref{fig: large_deviations}(b) has no discontinuity.
This is indicative of a second-order transition, with the rate function showing critical broadening, see Fig.~\ref{fig: large_deviations}(c), and 
a divergence in the susceptibility $\chi(s) = \theta''(s)$, see Fig.~\ref{fig: large_deviations}(d). 
For both models we can extract a transition point from the drop in either first or second cumulant. The top panel of Fig.~\ref{fig: large_deviations}(d) shows how the transition point scales with $L$ for both models (for a range of $c$ for the 2D East).
We are able to fit the data with the power laws $s_{c}(L) \sim L^{-2\alpha}$, as shown by the solid lines.
We find the exponents $\alpha \gtrsim 1$ for the 2D East and $\alpha \lesssim 1$ for the SSEP, see the inset to the top panel of Fig.~\ref{fig: large_deviations}(d).

\begin{figure*}[t]
    \centering
    \includegraphics[width=\linewidth]{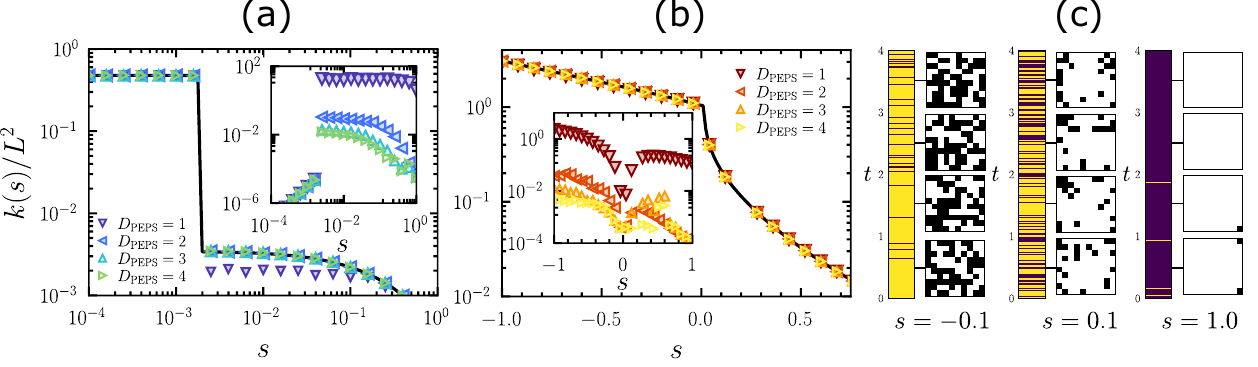}
    \caption{\textbf{Optimal sampling of trajectories.} (a) Average dynamical activity as a function of $s$ 
    for the 2D East model 
    from CTMC with importance sampling (symbols), for $c = 0.5$ and $D_{\rm PEPS} \in [1, 4]$.
    The trajectory times are chosen such that on average we expect 100 transitions per trajectory. 
    The solid black line shows the activity measured directly from the PEPS with $D = 4$ for comparison.
    Inset: variance in the time-integrated difference of escape rates, $\delta R^2$ (see main text). Each data point is calculated from $N_{\rm sp} \in [10^3, 10^4]$ trajectories.
    (b) Same but for the 2D SSEP on a $22 \times 22$ lattice.
    (c) Representative trajectories for the 2D SSEP for size $N = 10 \times 10$
    at three values of $s \neq 0$. 
    The bars on the left of each panel show the times when particle hops occur (yellow/bright lines). The snapshots on the right show the configurations at the marked times (black/white indicates a particle/hole). See also Refs.~\cite{Causer2022b_data, SM}.
    }
    \label{fig: optimal_sampling}
\end{figure*}

\textbf{\em Optimal sampling of rare trajectories from PEPS.-}
Sampling trajectories corresponding to the $s \neq 0$ phases is difficult as they are exponentially rare in system size and time. 
The optimal sampling dynamics at long times is given by the so-called {\em generalized Doob transform} \cite{Borkar2003,
Simon2009, Jack2010, Chetrite2015,Garrahan2016}, which maps the tilted generator into a true stochastic generator for the rare trajectories,
    $\mathbb{W}_{s}^{\rm Doob} = \mathbb{L}\left[ \mathbb{W}_{s} - \theta(s)\mathbb{I} \right]\mathbb{L}^{-1}$,
where $\mathbb{L}$ is the leading left eigenvector of $\mathbb{W}_{s}$ as a diagonal matrix.
This gives a new dynamics with the transition rates
\beq
    \Tilde{w}_{x\to y} = \frac{l_{s}(y)}{l_{s}(x)} e^{-s}w_{x\to y},
    \label{w_doob}
\eeq
with $l_{s}(x) = \braket{l_{s} | x}$. In 
$\mathbb{W}_{s}^{\rm Doob}$ the counting field $s$ appears as a physical control parameter, and running dynamics with rates \eqref{w_doob} gives trajectories at $s \neq 0$ on demand. While optimal, $\mathbb{W}_{s}^{\rm Doob}$ is difficult to construct in general as one needs the exact left leading eigenvector. However, we can exploit our PEPS approximation to estimate the rates \er{w_doob}, similar to Ref.~\cite{Causer2021} for 1D and MPS, see Ref.~\cite{SM} for more information. 

To obtain Eq.~\eqref{w_doob} for the transitions out of a state $x$ we calculate  $l_{s}(y)$ from the PEPS using a boundary dimension $\chi_{B} = D_{\rm PEPS}$ \cite{Ueda2005,Frias2021tnmh,Rams2021,Vieijra2021}, 
thus entailing a maximum cost $O(N D_{\rm PEPS}^6)$. 
If we neglect the time edges of trajectories, we can estimate an time-extensive observable by importance sampling
\beq
    \braket{O}_{s} \approx \frac{\sum_{\alpha_{t}} O(\alpha_{t})g(\alpha_{t})} {\sum_{\alpha_{t}} g(\alpha_{t})},
    \label{Oref}
\eeq
where $\alpha_{t}$ denotes a trajectory generated with \er{w_doob} (the {\em reference dynamics}), and $O(\alpha_{t})$ is the trajectory observable.
The re-weighting factor $g(\alpha_{t})$ is 
\beq
    g(\omega_{t}) = e^{-\int_{0}^{t} dt' R(t') - \Tilde{R}(t')},
    \label{gref}
\eeq
where $R(t')$ and $\Tilde{R}(t')$ are the escape rates of the system at time $t'$ in the original dynamics and the approximate Doob dynamics, respectively.
Notice that with a large enough number of trajectories, \er{Oref} can be used to correct on the imperfections in the reference dynamics due to an imperfect PEPS approximation.

Figures \ref{fig: optimal_sampling} show results from our sampling algorithm for the 2D East with $c = 0.5$ and the 2D SSEP, both for system sizes $N=22\times 22$. The average dynamical activity measured in trajectories (symbols) [with umbrella sampling Eqs.(\ref{Oref},\ref{gref})] coincides with that obtained directly from the PEPS (solid line), except for $D_{\rm PEPS} = 1$ for the East model.
The accuracy of our dynamics is quantified by the variance of the time integrated difference in escape rates, cf.~\er{gref}, which vanishes for the exact Doob rates.
We show this for each $D$ in the insets of Figs.~\ref{fig: optimal_sampling}: increasing the $D_{\rm PEPS}$ consistently reduces the variance, indicating a better sampling dynamics and less need for importance sampling.

The PEPS approximation to the leading eigenvalue gives us direct access to the long time-averaged properties of the dynamics. 
However, the broader range of dynamical information---such as temporal correlations---can only be obtained through the simulation of the dynamics in rare dynamical regimes.
The ability to exploiting the PEPS to define an efficient trajectory sampling scheme for the biased dynamics allows us to characterize the finite-time atypical behaviour beyond what is directly encoded in the PEPS approximation.
Figure \ref{fig: optimal_sampling}(c) illustrates this by showing sample trajectories for the SSEP at $s\neq0$: in the active phase ($s=-0.1$ panel), particle hops are plentiful, as shown in the bar on the left of the panel, and the configurations visited in the trajectory have a finite density and show no appreciable clustering; in contrast in the inactive phase  
($s=1.0$ panel), hops are scarce and the configurations have very low density of particles; near the transition ($s=0.1$ panel), configurations show more pronounced density fluctuations, related to the critical nature of the transition.

\textbf{\em Discussion.-}
We have shown that the dynamical LDs of two-dimensional stochastic models can be studied efficiently with PEPS, including the quasi-optimal sampling of atypical trajectories. 
Compared to more standard methods \cite{Giardina2006,Cerou2007,Lecomte2007b,Nemoto2016,Ray2018,Klymko2018,Ferre2018}, PEPS offer the advantage of a computational cost that scales only polynomially in the system size and the tensor dimensions $D_{\rm PEPS}$. Furthermore, the algorithms produce an explicit ansatz for the leading eigenvector encoding the LDs, which allows to extract all (local) observables in the biased trajectory ensemble and, as shown above, to devise a near-to-optimal sampling dynamics. Additionally, because PEPS form a complete family, by increasing the bond dimension the quality of the solution can be systematically improved, a property that can also be utilized to estimate the error of the approximation. Our results show that the PEPS ansatz is well suited for these problems, as a moderate bond dimension suffices to explore large systems.

We showed here that both the 2D East model and the 2D SSEP have active-inactive trajectory transitions, of the first-order and second-order, respectively. Our work adds to the continuously expanding application 
\cite{Gorissen2009, Gorissen2012, Gorissen2012a, Banuls2019, Helms2019, Causer2020, Helms2020, Causer2022a, Gu2022, Causer2022, Strand2022, Strand2022a,Garrahan2022}
of tensor network methods to study the dynamical fluctuations in classical stochastic systems. 

There are several interesting avenues to pursue building on this work. One is to integrate 2D trajectory sampling via TNs with a method such as transition path sampling \cite{Bolhuis2002} for investigating statistics of fluctuations at {\em finite times}, cf. Ref~\cite{Causer2022, Tang2022}.
While the current implementations with PEPS are too demanding to reasonably incorporate transition path sampling, tree tensor networks \cite{Shi2006} are a promising alternative that could allow to reliably investigate finite time scaling. We hope to report on this is the near future.

\acknowledgements 
\textbf{\em Acknowledgements.-}
We acknowledge financial support from EPSRC Grant no.\ EP/R04421X/1 and the Leverhulme Trust Grant No. RPG-2018-181. 
M.C.B.\ acknowledges support from Deutsche Forschungsgemeinschaft (DFG, German Research Foundation) under Germany's Excellence Strategy -- EXC-2111 -- 390814868. 
Calculations were performed using the Sulis Tier 2 HPC platform hosted by the Scientific Computing Research Technology Platform at the University of Warwick. Sulis is funded by EPSRC Grant EP/T022108/1 and the HPC Midlands+ consortium.
We acknowledge access to the University of Nottingham Augusta HPC service. 
Example code used to generate the data can be found at \url{https://github.com/lcauser/2d-optimal-sampling}.
Data is available at Ref.~\cite{Causer2022b_data}.
 
\bibliographystyle{apsrev4-2}
%\bibliography{bibliography}
\nocite{Schuch2007complexity, Corboz2010, Phien2015, Suzuki1985}

\newpage
\onecolumngrid
\begin{appendix}
\section{Supplemental Material} 
\setcounter{figure}{0}
\setcounter{equation}{0}
\makeatletter 
\renewcommand{\thefigure}{S\@arabic\c@figure}
\renewcommand{\theequation}{S\@arabic\c@equation}
\makeatother

\section{Stochastic dynamics}
We consider systems that live on a two-dimensional square lattice of size $N = L \times L$, where each site can take the binary values $n_{\bm k} = 0$ or $1$, and $\bm{k} = (k_{x}, k_{y})$ denotes the position of the lattice sites, $k_{x}, k_{y} = 1\cdots L$.
The system evolves under continuous-time Markov dynamics, defined by the transition rates $w_{x\to y}$ from configuration $x$ to $y$.
The average dynamics of the system can be encoded by a probability distribution $P_{x}(t)$, which describes the probability for the system to be in some configuration $x$ at time $t$.
This can be compactly written as a vector of probabilities, $\ket{P(t)} = \sum_{x} P_{x}(t) \ket{x}$, where $\sum_{x} P_{x}(t) = 1$.
The evolution of the probability distribution is given by the master equation,
\beq
    \frac{d}{dt} P_{x}(t) = \sum_{y \neq x} w_{y\to x}P_{t}(y) - R_{x} P_{t}(x),
\eeq
where $R_{x} = \sum_{y \neq x} w_{x \to y}$ is the escape rate out of the configuration $x$.
It is convenient to write the master equation in terms of a Markov generator,
\beq
    \mathbb{W} = \sum_{x, y\neq x} w_{x\to y}\ket{y}\bra{x} - \sum_{x} R_{x} \ket{x}\bra{x},
\eeq
which yields $\partial_{t} \ket{P(t)} = \mathbb{W}\ket{P(t)}$.
The Markov generator $\mathbb{W}$ conserves probability, $\bra{-} \mathbb{W} = 0$, with the {\em flat state} $\bra{-} = \sum_{x} \bra{x}$, 
and has maximal eigenvalue zero.
%and zero the maximal eigenvalue of $\mathbb{W}$.

For the models considered here, the Markov generator can easily be written in terms of local operators.
The first model we consider is the 2D East model, with the Markov generator
\beq
    \mathbb{W}^{\rm East} = \sum_{\bm k} P_{\bm k} \Big[ c\left(\sigma_{\bm k}^{+}  - (1-n_{\bm k})\right)
     + (1-c)\left(\sigma_{\bm k}^{-} - n_{\bm k}\right)\Big],
     \label{W_east}
\eeq
where $\sigma^{\pm}_{\bm k}$ are the Pauli raising/lowering operators at site ${\bm k}$, $c\in(0, 1/2]$ parameterizes the transition rates, and the kinetic constraint is
$P_{(k_x, k_y)} = n_{(k_{x}-1,\,k_{y})} + n_{(k_{x},\,k_{y}-1)}$. 
The second model we consider is the 2D symmetric simple exclusion process (SSEP),
\beq
    \mathbb{W}^{\rm SSEP} = \sum_{\left<{\bm k}, {\bm l}\right>} \Big[
    \sigma_{\bm k}^{+}\sigma_{\bm l}^{-} - (1-n_{\bm k})n_{\bm l}
    +\sigma_{\bm k}^{-}\sigma_{\bm l}^{+}
      - n_{\bm k}(1-n_{\bm l}) \Big]
     + \frac{1}{2}\sum_{{\bm k}\in\partial}\Big[
    \sigma^{x}_{\bm k} - 1\Big],
    \label{W_sep}
\eeq
where $\left<{\bm k}, {\bm l}\right>$ denotes a pair of nearest neighbours, $\sigma_{i}^{x} = \sigma_{i}^{+} + \sigma_{i}^{-}$, and $\partial$ denotes the boundary of the lattice.
See the main text for a description of each.

\subsection{Mapping the tilted generator onto a Hermitian operator}
The long time statistics for the models are encoded in the leading eigenvalue $\theta(s)$ and eigenvectors of the tilted Markov generator, $\mathbb{W}_{s}$, retrieved by multiplying the off-diagonal components of \ers{W_east}{W_sep} by $e^{-s}$,
\begin{gather}
    \bra{l_{s}} \mathbb{W}_{s} = \theta(s) \bra{l_{s}} \mathbb{W}_{s},
    \\
    \mathbb{W}_{s} \ket{r_{s}} = \theta(s) \mathbb{W}_{s} \ket{r_{s}}.
\end{gather}
The probability distribution over the configuration space under these biased dynamics then behaves as
%goes
\beq
    P(x) \asymp \frac{\braket{l_{s} | x}\braket{x | r_{s}}} {\braket{l_{s} | r_{s}}}
\eeq
for sufficiently large times.
The tilted Markov generator provides an efficient way to estimate the time-averaged dynamical properties by directly targeting its leading eigenvectors.
However, we can exploit the fact that each of the models considered here obeys detailed balance. 
We define $\mathbb{P}$ as the diagonal matrix whose elements are the square roots of the steady state probability of $\mathbb{W}$.
Then we can use the similarity transformation
$
    \mathbb{H}_{s} = \mathbb{P}^{-1}\mathbb{W}_{s}\mathbb{P},
    \label{Hs}
$
to define the Hermitian matrix $\mathbb{H}_{s}$, with maximal eigenvalue and associated eigenvector
\beq
    \mathbb{H}_{s}\ket{\psi_{s}} = \theta(s)\ket{\psi_{s}},
\eeq
where $\ket{\psi_{s}}$ is related to the original eigenvectors by $\ket{r_{s}} = \mathbb{P}\ket{\psi_{s}}$ and $\bra{l_{s}} = \bra{\psi_{s}} \mathbb{P}^{-1}$.
The models considered here allow for a simple representation as a Hermitian matrix,
\begin{gather}
    \mathbb{H}^{\rm East}_{s} = \sum_{\bm k} P_{\bm k} \Big[ 
    e^{-s}\sqrt{c(1-c)} \sigma_{i}^{x} - c(1-n_{\bm k}) - (1-c)n_{\bm k}\Big],
    \label{H_east}
    \\
    \mathbb{H}^{\rm SSEP}_{s} = \sum_{\left<{\bm k}, {\bm l}\right>} \Big[
    e^{-s}\sigma_{\bm k}^{+}\sigma_{\bm l}^{-} - (1-n_{\bm k})n_{\bm l}
    + e^{-s}\sigma_{\bm k}^{-}\sigma_{\bm l}^{+}
      - n_{\bm k}(1-n_{\bm l}) \Big]
     + \frac{1}{2}\sum_{{\bm k}\in\partial}\Big[
    e^{-s}\sigma^{x}_{\bm k} - 1\Big].
    \label{H_sep}
\end{gather}

This representation is convenient due to the fact its expectation with any vector $\ket{\psi}$ is bounded by the maximal eigenvalue through the Rayleigh-Ritz variational principle,
\beq
    \frac{\braket{\psi | \mathbb{H}_{s} | \psi}}{\braket{\psi | \psi}} \leq \theta(s).
\eeq
Furthermore, using this Hermitian matrix means that we only need to determine one eigenvector.
Notice that in this representation, the leading eigenvector encodes the {\em probability amplitudes} of each configuration, $\ket{\psi_{s}} = \sum_{x} \psi(x) \ket{x}$, with $|\psi(x)|^2 = P(x)$.

\section{Projected entangled-pair states}
The long-time dynamics of deformed Markov generators can be encoded by the probability amplitude, $\psi({\bm n})$, with configurations ${\bm n} = (n_{{\bm k}_{1}}, n_{{\bm k}_{2}}, \cdots, n_{{\bm k}_{N}})$.
This can be written more compactly as a vector of probability amplitudes, $\ket{\psi} = \sum_{\bm n} \psi({\bm n}) \ket{\bm n}$.
This vector has the size $2^{N}$ for $N$ lattice sites, and quickly becomes intractable to store.
%However, it can often be the case that many of vector elements $\psi({\bm n})$ share {\em mutual information}, allowing for a more efficient representation.
However, it is often the case that the components $\psi({\bm n})$ are not independent of each other, such that it is possible to efficiently approximate  the vector
%Moreover, these states can often be approximated 
by an object with a smaller dimensionality (described by a number of parameters less than $d^N$). 
This realisation is at the heart of tensor network (TN) approximations.
A TN representation of $\psi({\bm n})$ amounts to decomposing the $N$-dimensional object into a network of many smaller tensors, connected along additional virtual dimensions, 
which are contracted (i.e. multiplied and summed over) to retrieve the original global tensor.
%TNs allow for a general representation of $\psi({\bm n})$ by composing the tensorial object $\psi({\bm n})$ (it is tensorial because its value relies on a specification of each subsystem, $n_{\bm{k}}$) 
%into many smaller tensors which can be contracted (multiplied and summed over).
%Typically, each subsystem is assigned its own tensor, with the introduction of virtual dimensions which connect it to other subsystems. 

\begin{figure}[t]
    \centering
    \includegraphics[width=\linewidth]{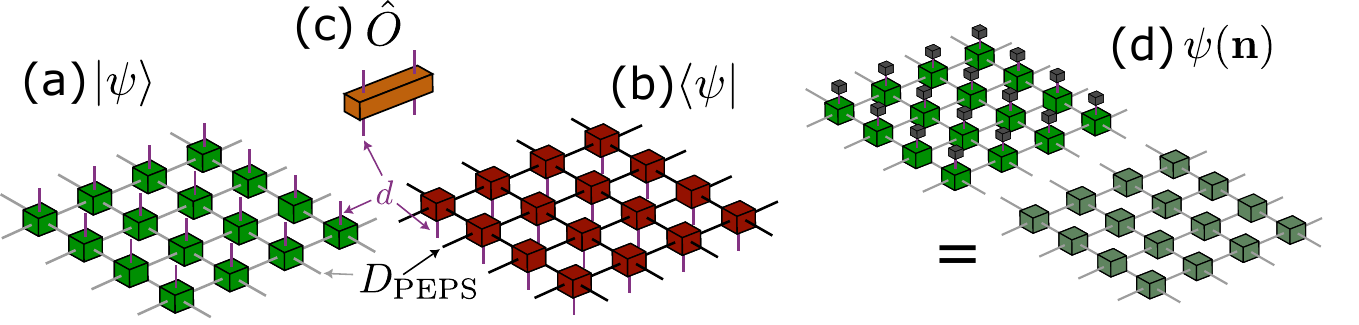}
    \caption{
        {\bf PEPS.}
        (a) The vector $\ket{\psi}$ as a PEPS. The green cubes represent tensors within the PEPS, the (open) purple lines are their physical dimensions, and the (closed) grey lines are the virtual dimensions between tensors.
        (b) The same for its conjugate, $\bra{\psi}$. The conjugate tensors are represented by red cubes and black lines.
        (c) The two-body operator, $\hat{O}$, as a tensor.
        (d) The probability amplitude $\psi(\bm{n})$ can be retrieved by specifying the local dimension of each tensor, as shown by the grey cubes.
        This results in a network of rank-4 tensors, which give $\psi(\bm{n})$ when contracted.
    }
    \label{fig: sm_peps}
\end{figure}

For the case of a 2D square lattice with $N = L\times L$ sites, the most natural TN ansatz is the {\em projected entangled-pair state} (PEPS).
%Each subsystem 
In this ansatz, each system (lattice site) is assigned its own rank-5 tensor, $A_{{\bm k}_{j}}^{d_{j}}$, where ${\bm k}_{j}$ denotes the position of system $j$, and $d_{j}$ the state of the system.
One of the dimensions corresponds to the physical dimension of the subsystem with size $d$, and the other four dimensions are virtual ones which connect the tensor to the tensors of the four neighbouring lattice sites.
These virtual dimensions are of size $D_{\rm PEPS}$ (often referred to as the {\em bond dimension}), which controls the amount of mutual information shared between the lattice sites.
It follows that each tensor is parameterized by $dD_{\rm PEPS}^{4}$ parameters, with a maximum of $N_{p} = NdD_{\rm PEPS}^{4}$ parameters for the whole PEPS.
By specifying the local configuration of each site, $n_{\bm k}$, and contracting over the network, one is able to determine $\psi({\bm n})$,

\beq
    \psi({\bm n}) = \mathcal{F} \left(A_{{\bm k}_{1}}^{n_{\bm{k}_{1}}} A_{{\bm k}_{2}}^{n_{\bm{k}_{2}}} \cdots A_{{\bm k}_{N}}^{n_{\bm{k}_{N}}} \right),
\eeq
where $\mathcal{F}$ is a function which represents the contraction over all virtual bond dimensions.
It is convenient to represent the TN pictorially, as illustrated for $\ket{\psi}$ in Fig.~\ref{fig: sm_peps}(a).
Notice that we refer to this as $\ket{\psi}$, as none of the physical dimensions are specified. 
The green cubes correspond to the tensors, the grey lines represent the virtual bond dimensions to be contracted over, and the pink lines represent the physical dimensions.
Similarly, we show the conjugate, $\bra{\psi}$, in Fig.~\ref{fig: sm_peps}(b).
It is also possible to represent local operators as a tensor, such as the two-body operators in the tilted generators.
Local two-body operators can be represented by a rank-4 tensor, as shown in Fig.~\ref{fig: sm_peps}(c).
Figure \ref{fig: sm_peps}(d) shows how the probability amplitude $\psi(\bm{n})$ can be retrieved from the PEPS by specifying the local dimension for each system.
This reduces the PEPS to a network of rank-4 tensors, which give $\psi(\bm{n})$ when contracted.

\subsection{Contracting PEPS}
\begin{figure}[t]
    \centering
    \includegraphics[width=0.6\linewidth]{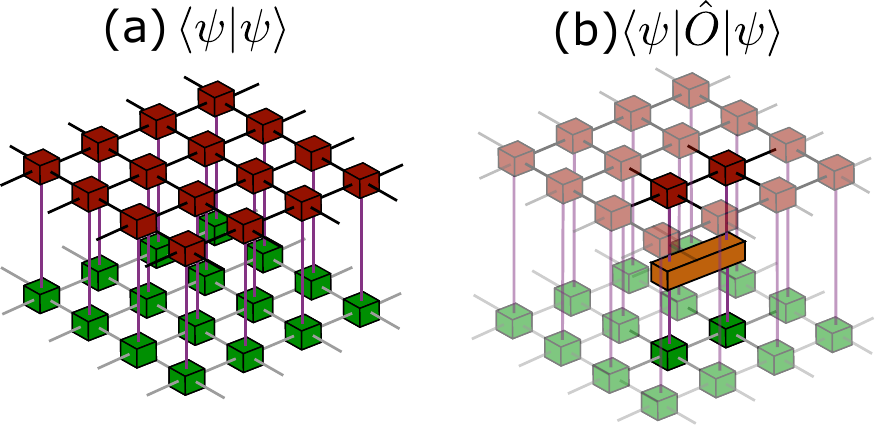}
    \caption{
        {\bf PEPS networks.}
        (a) The inner product $\braket{\psi | \psi}$ and (b) the expectation value $\braket{\psi | \hat{O} | \psi}$ as TNs.
        For visual convenience, the operator and the tensors which it is connected to are show in full colour, while the remaining tensors are semitransparent.
        %opaque.
    }
    \label{fig: sm_peps_networks}
\end{figure}

PEPS allow for an efficient representation of probability amplitude vectors for large system sizes.
However, to perform any tractable calculations, such as calculating the expectation of observables, it is necessary to have a way to efficiently contract the networks.
Figures \ref{fig: sm_peps_networks}(a, b) show the networks which must be contracted to determine the inner products $\braket{\psi | \psi}$ and $\braket{\psi | \hat{O} | \psi}$ respectively, for some two-body operator $\hat{O}$.
In general, contracting exactly such a two-dimensional TN is an intractable problem~\cite{Schuch2007complexity}, as any exact contraction strategy has a cost that scales at least exponentially in $L$.
%For some general system size, there is no efficient way to calculate these exactly.
%The most optimal way to contract this exactly is one row (or column) of lattice sites at a time.
%However, this is exponentially costly in $L$.
%Instead, 
For the ansatz to be of practical use, one needs ways to approximately (but precisely) contract the TN with a tractable cost.

\begin{figure}[t]
    \centering
    \includegraphics[width=0.8\linewidth]{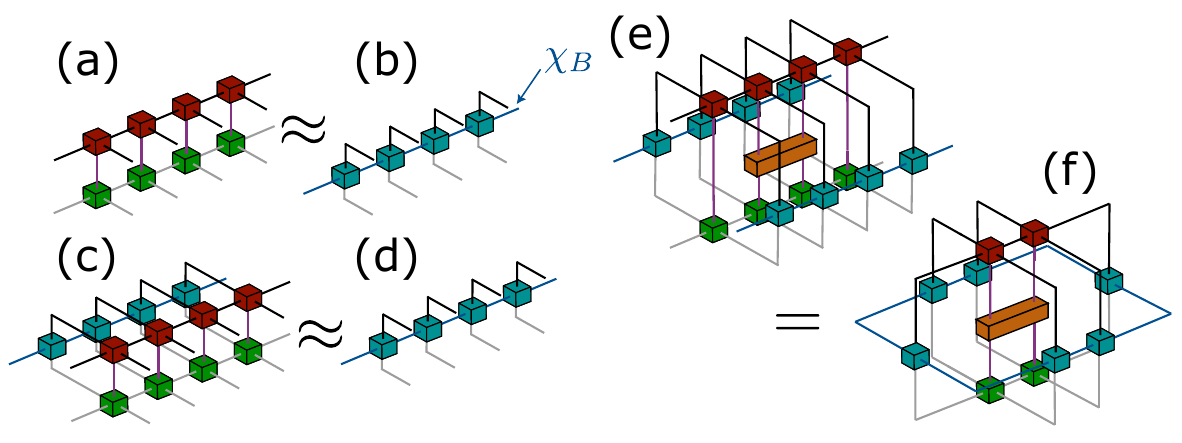}
    \caption{
        {\bf Contracting PEPS networks.}
        (a) The edge of a PEPS network can be approximated by (b) a boundary MPS with the auxiliary bond dimension $\chi_{B}$.
        (c) The boundary MPS can be contracted with subsequent layers within the PEPS network, and approximated by (d) another boundary MPS also with the auxiliary bond dimension $\chi_{B}$.
        By contracting from two opposite edges of the PEPS network, we have the reduced network shown in (e).
        One can then exactly contract from both edges of the reduced network to receive the network shown in (f). This can be exactly contracted to give the approximation of $\braket{\psi | \hat{O} | \psi}$.
    }
    \label{fig: sm_peps_contract}
\end{figure}

A popular approach, and the one we will take here, is the boundary matrix product state (MPS) scheme \cite{Verstraete2004, Verstraete2008}.
This involves contracting from the edge of the TN, one row (column) at a time, and approximating the result by an MPS (in this case, a collection of rank-4 tensors), 
%with the bond dimension $\chi_{\rm B}$ controlling 
whose bond dimension $\chi_{\rm B}$ controls the accuracy of the approximation.
We demonstrate this process in Figs.~\ref{fig: sm_peps_contract}(a-d).
The tensors at the boundary can be contracted to form the first boundary MPS, see Fig.~\ref{fig: sm_peps_contract}(a).
This boundary MPS is then contracted with the subsequent layer, and approximated by another boundary MPS, again with bond dimension $\chi_{\rm B}$, see Fig.~\ref{fig: sm_peps_contract}(a). 
Applying the procedure from two opposing edges of the network up to the rows (or columns) of lattice sites neighbouring the two-body operator, one is able to approximate the expectation value depicted in Fig.~\ref{fig: sm_peps_networks}(b) by the one shown in 
Fig.~\ref{fig: sm_peps_contract}(e).
This network can then be contracted from the edges exactly, resulting in Fig.~\ref{fig: sm_peps_contract}(f), which can be easily and exactly contracted to determine the approximation of $\braket{\psi | \hat{O} | \psi}$.
Contracting the networks in Fig.~\ref{fig: sm_peps_contract} scales as $O(\chi_{B}^3 D_{\rm PEPS}^{4} + \chi_{B}^2 D_{\rm PEPS}^{6})$.

\begin{figure}[t]
    \centering
    \includegraphics[width=0.5\linewidth]{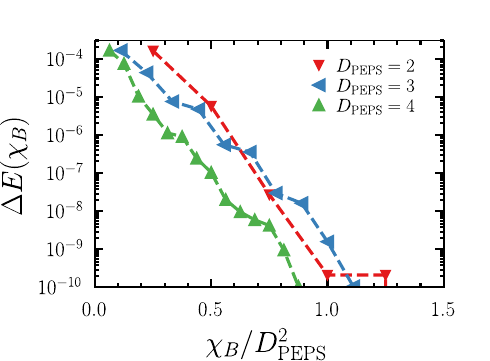}
    \caption{
        {\bf Accuracy of the boundary approximation.}
        We show the relative difference in the expectation value, $\Delta E(\chi_{B})$, when a boundary MPS bond dimension $\chi_{B}$ is used, compared to $\chi_{B} = 128$.
        Results are for the 2D East model with $N = 10\times 10$, $c = 0.5$ and $s = -1.0$.
    }
    \label{fig: sm_peps_boundary}
\end{figure}

A heuristic choice for $\chi_{B}$ is $\chi_{B} \sim O(D^2)$, in which case gives the total complexity $O(D^{10})$.
Figure \ref{fig: sm_peps_boundary} demonstrates the role of the bond dimension of the boundary MPS, $\chi_{B}$.
We optimize the MPS using the simple update described below, and calculate its expectation value with respect to $\mathbb{H}_{s}$, $E(\chi_{B}) = \braket{\psi_{s} | \mathbb{H}_{s} | \psi_{s}}$, measured with the bond dimension $\chi_{B}$.
We compare the measurement to that using $\chi = 128$, which can be considered to be quasi-exact,
\beq
    \Delta E(\chi_{B}) = \left|\frac{E(\chi_{B}) - E(128)}{E(128)}\right|.
\eeq
Indeed, the results confirm that $\chi_{B} \sim O(D^2)$ is a reasonable choice for approximating the environment.
We use the bond dimension $\chi = 50$ when checking for convergence in observables during the optimization procedure, and $\chi = 200$ when taking the final measurements to ensure convergence in the boundary approximation.

\subsection{Time evolution}
To find a PEPS approximation for the leading eigenvector of the tilted generator, one needs to find a suitable optimization procedure.
There are many approaches to optimizing TNs, but the one taken here will be to employ {\em time evolution} (often referred to imaginary time evolution for quantum many-body systems).
The main idea is to project some probability amplitude vector onto the leading eigenvector of the tilted generator $\mathbb{H}_{s}$ by applying the time propagator operator, $U(\delta) = e^{\delta \mathbb{H}_{s}}$, to $\ket{\psi}$ until convergence is met, $\ket{\psi_{s}} = \lim_{t \to \infty} U(t)\ket{\psi}$ (up to normalization).
In practice, the complete time propagation operator is difficult to compute, as it requires the matrix exponentiation of the tilted generator on the complete state space.
However, for small $\delta \ll 1$, we can approximate the time propagation by a sequence of {\em Trotter gates},
\beq
    U(\delta) \approx \prod_{\left\langle {\bm k}, {\bm l} \right\rangle} U_{{\bm k}, {\bm l}}(\delta),
\eeq 
where $U_{{\bm k}, {\bm l}}(\delta) = e^{\delta H_{{\bm k}, {\bm l}}}$, and $H_{{\bm k}, {\bm l}}$ are the terms in $\mathbb{H}_{s}$ which act only on the lattice sites ${\bm k}$ and ${\bm l}$.
This approximation is often referred to as a first-order Trotter decomposition, with each set of gates having an error of $O(\delta^2)$.
This is the approach we take.
However, it is possible to improve the accuracy by using higher order Trotter decompositions ~\cite{Suzuki1985}.

\begin{figure}[t]
    \centering
    \includegraphics[width=\linewidth]{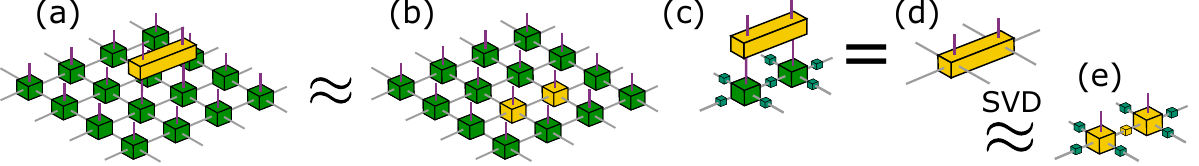}
    \caption{
        {\bf Time evolution.}
        (a) The PEPS $\ket{\psi}$ can be updated at a local neighbouring pair of sites through the application of a Trotter gate (shown by the yellow cuboid).
        (b) This can be approximated by a PEPS with same bond dimension.
        (c) The updated tensors can be retrieved using the SU.
        The small turqouise cubes are the ``$\lambda$-matrices'' retrieved from SVDs of the neighbouring tensors (see e.g. Ref.~\cite{Jiang2008}).
        (d) First, we contract the lattice site tensors and the surrounding $\lambda$-matrices with the Trotter gate.
        (e) Then, through an SVD, we can restore the PEPS manifold.
        The $\lambda$-matrices outside of the pair of tensors are restored to their original values, while the $\lambda$-matrix between the two tensors is updated.
    }
    \label{fig: sm_peps_update}
\end{figure}

Each gate can be individually applied to the PEPS, see Fig.~\ref{fig: sm_peps_update}(a).
The goal is to approximate the application of the gate to PEPS by another PEPS with the same bond dimension, see Fig.~\ref{fig: sm_peps_update}(b).
Naively contracting the gate moves the PEPS away from the PEPS manifold, as shown in Figs.~\ref{fig: sm_peps_update}(c, d).
In order to restore the PEPS, we need an update scheme which restores the form of the original two tensors from the PEPS used in the contraction.
Two popular approaches to achieving this task are the Simple Update (SU) \cite{Jiang2008} and the Full Update (FU) \cite{Corboz2010, Lubasch2014,Phien2015}.
In general, to find the optimal truncation, which optimizes the overlap of the new PEPS with the untruncated TN, the environment of the pair of tensors 
(i.e. the contraction of the remaining PEPS tensors) needs to be taken into account. 
This is however a costly computation, and the SU includes only a primitive, but computationally inexpensive approximation of it as a product~\cite{Lubasch2014unifying},
and performs a truncated singular value decomposition (SVD) to split the resulting tensor, see Figs.~\ref{fig: sm_peps_update}(d, e). 
This results in a simple algorithm, with computational cost scaling as $O(D_{\rm PEPS}^5)$.
\footnote{As presented here, the complexity is $O(D_{\rm PEPS}^6)$, but can be made cheaper by considering the reduced tensor method \cite{Lubasch2014}.}.

In contrast, the FU takes into account the full environment, but requires the approximate contraction of the complete TN, as illustrated in Figs.~\ref{fig: sm_peps_networks} and \ref{fig: sm_peps_contract}.
This update has a greater accuracy, but with a much steeper scaling of $O(\chi_{B}^3 D_{\rm PEPS}^{4} + \chi_{B}^2 D_{\rm PEPS}^{6})$. 

\begin{figure}[t]
     \centering
     \includegraphics[width=0.5\linewidth]{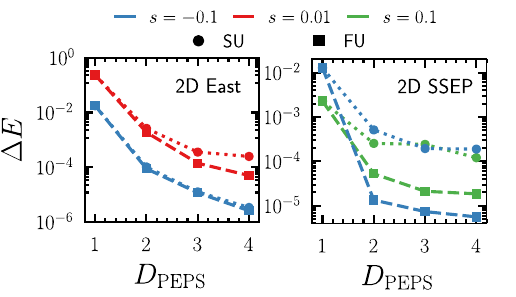}
     \caption{\textbf{Optimization of PEPS.} 
     The error in the measured expectation value for the SU and the FU compared to the high accuracy 2D DMRG (with an MPS bond dimension up to $D_{\rm MPS} = 1024$) for various values of $s$ and a $10\times10$ lattice. 
     The left panel shows the 2D East model with $c = 0.5$, and the right panel shows the 2D SSEP.
     The PEPS environment in the FU uses a boundary dimension $\chi_{B} = 4D^2$ for the East and $\chi_{B} = 6D^2$ for the SSEP.
     }
     \label{fig: sm_peps_error}
 \end{figure}

In what follows, we demonstrate the effectiveness of both methods for the system sizes $N = 10 \times 10$.
For both approaches, we use an update schedule which reduces the time step from in the range $\delta \in [10^{-3}, 10^{-1}]$.
After many iterations of time evolution, we estimate the expectation value of the state, $E = \braket{\psi | \mathbb{H}_{s} | \psi}$.
This process is repeated until we find convergence.
Figures \ref{fig: sm_peps_error}(a, b) show the difference in the expectation value, $\Delta E = (E - E_{\rm DMRG}) / E_{\rm DMRG}$, with respect to the quasi-exact results of 2D density matrix renormalization group (DMRG) for the 2D East and 2D SSEP respectively, and various values of $s$.
The circles show the results of the SU, and the squares show the results of the FU.
While it is clear in most instances the FU can provide significant improvements on the SU, it is worth noting that even the SU provides precise results for bond dimension $D = 4$, with errors $\delta E \lesssim O(10^{-3})$.
We find these errors to be sufficiently small, and thus proceed using only the SU to allow us to reach large system sizes.

 \section{Optimal sampling of dynamics}
 The long time  statistics of dynamical observables are encoded in the deformed Markov generator $\mathbb{W}_{s}$.
 While in principle this object can be used to generate the trajectories which correspond to the statistics, it is difficult to do in practice due to the unnormalized nature of $\mathbb{W}_{s}$.
 That is, $\bra{-}\mathbb{W}_{s} \neq 0$, and in general, the leading eigenvalue $\theta(s) \neq 0$.
 In the long time limit, we are able to overcome this difficulty using the so-called {\em Doob dynamics}, which maps the tilted generator onto a proper stochastic dynamics through the transformation
    $\mathbb{W}_{s}^{\rm Doob} = \mathbb{L} \left[ \mathbb{W}_{s} - \theta(s) \mathbb{I} \right] \mathbb{L}^{-1}$,
where $\mathbb{L} = \sum_{x} l_{s}(x) \ket{x}\bra{x}$ is the left eigenvector as a diagonal matrix, $l_{s}(x) = \braket{l_{s} | x}$. 
It is simple to check that the flat state $\bra{-}$ is an eigenvector of $\mathbb{W}_{s}^{\rm Doob}$ with the maximal eigenvalue zero.
The Doob dynamics has the transition rates
\beq
    \Tilde{w}_{x\to y} = \frac{l_{s}(y)}{l_{s}(x)} e^{-s}w_{x\to y},
    \label{w_doob}
\eeq
and escape rates $\Tilde{R}_{x} = R_{x} + \theta(s)$.

While the Doob dynamics provides an efficient way to simulate the biased dynamics at long times, it is dependent on the fact that one has access to the leading eigenvector of the tilted generator. 
Using the PEPS optimization methods described above, we are able to estimate the left eigenvector $\bra{l_{s}} \approx \bra{\psi_{s}} \mathbb{P}^{-1}$, where $\bra{\psi_{s}}$ is our approximation to the leading eigenvector of $\mathbb{H}_{s}$.
Note that retrieving $\bra{l_{s}}$ as a PEPS is simple due to the fact that $\mathbb{P}$ acts locally.
This allows us to implement an efficient sampling algorithm which can estimate the dynamics \er{w_doob} by sampling $l_{s}(x)$ directly form our PEPS.

\begin{figure}[t]
    \centering
    \includegraphics[width=\linewidth]{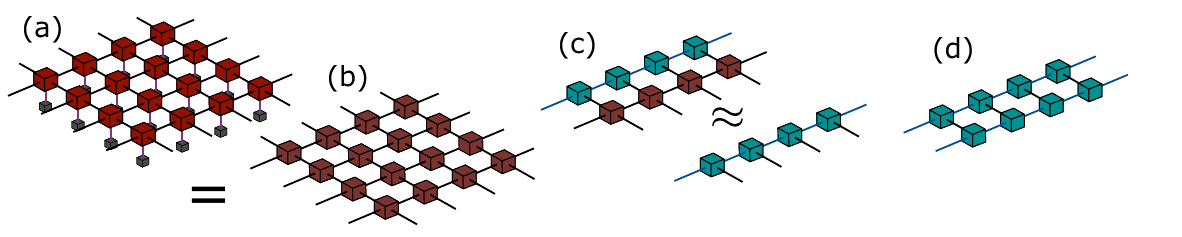}
    \caption{\textbf{Optimal sampling from PEPS.} 
    (a) The component $l_{s}(x) = \braket{l_{s} | x}$ can be extracted from the PEPS.
    (b) The physical dimensions can be contracted to give a PEPS composed of rank-4 tensors.
    (c) As is the case in Fig.~\ref{fig: sm_peps_contract}, the contraction of the PEPS can be estimated through a boundary MPS, this time composed of rank-3 tensors and bond dimension $\chi_{B}$.
    (d) By contracting from two opposing edges of the network, we can reduce approximate the contraction as an MPS-MPS product.
    }
    \label{fig: sm_peps_sampling}
\end{figure}

Extracting $l_{s}(x)$ from the PEPS is done similarly to estimating the contraction of the PEPS networks in Fig.~\ref{fig: sm_peps_networks}.
The first thing to notice is that we can reduce the PEPS to a network of rank-4 tensor by specifying 
the value of the local index of each tensor, which is defined by the configuration $x$ (that is, the configuration $x$ specifies each local $n_{\bm k}$), see Figs.~\ref{fig: sm_peps_sampling}(a, b).
The values $l_{s}(x)$ are then retrieved by contracting the network.
As was done for the networks in Fig.~\ref{fig: sm_peps_contract}, we can estimate the exact contraction of the network using the boundary MPS method, with some bond dimension $\chi_{B}$.
However, this time the boundary MPS are composed of rank-3 tensors, and a heuristic choice for the bond dimension is $\chi_{B} \sim O(D_{\rm PEPS})$, see Fig.~\ref{fig: sm_peps_sampling}(c).
By contracting from two opposing edges of the network, we can then estimate $l_{s}(x)$ through the exactly contractable MPS-MPS product, see Fig.~\ref{fig: sm_peps_contract}(d).

Unlike the networks in Fig.~\ref{fig: sm_peps_networks}---which are composed of two PEPS layers---the sampling of $l_{s}(x)$ only requires us to contract over a single PEPS. 
This leads to a significant reduction in computational cost, with each calculation of $l_{s}(x)$ only costing $O(ND_{\rm PEPS}^{6})$.
At each Monte Carlo step in the stochastic simulation algorithm, we need to calculate $l_{s}(x)$ for a maximum of $N$ configurations, and thus the cost is $O(N^{2} D_{\rm PEPS}^{6})$.
However, by recycling our partial contractions when calculating each $l_{s}(x)$, it is possible to reduce this to a cost of $O(N D_{\rm PEPS}^{6})$ for each Monte Carlo step, see Refs.~\cite{Phien2015, Causer2021}.
While our approach provides a way to nearly optimally sample the Doob dynamics, the PEPS used in the sampling is only approximate. 
These errors must be accounted for through the use of umbrella sampling, see the main text and Ref.~\cite{Causer2021} for further details.

\begin{figure}[t]
    \centering
    \includegraphics[width=0.5\linewidth]{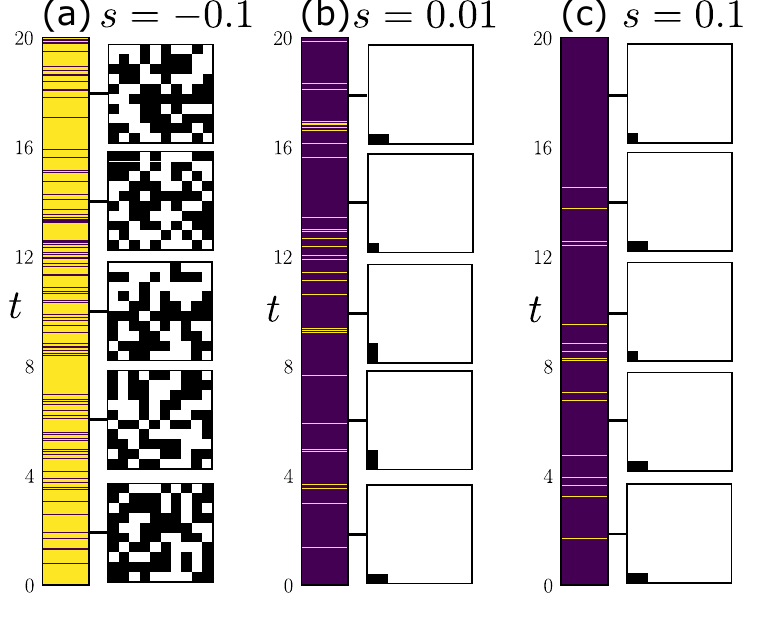}
    \caption{\textbf{Representative trajectories for the 2D East.} 
    Trajectories sampled from the approximate Doob dynamics for $L = 10$, $c = 0.5$, and (a) $s = -0.1$, (b) $s = 0.01$ and (c) $s = 0.1$.
    The vertical bars show the times when jumps occur (yellow/bright lines).
    The snapshots show the configurations at the marked times (black/white indicates a occupied/unoccupied).
    }
    \label{fig: sm_traj_east}
\end{figure}

\begin{figure}[t]
    \centering
    \includegraphics[width=0.5\linewidth]{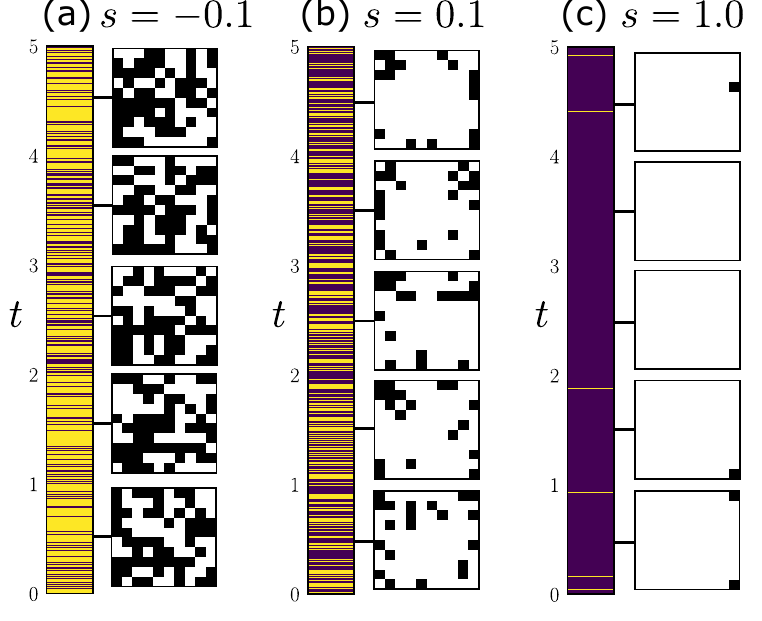}
    \caption{\textbf{Representative trajectories for the 2D SSEP.} 
    Trajectories sampled from the approximate Doob dynamics for $L = 10$ and (a) $s = -0.1$, (b) $s = 0.1$ and (c) $s = 1.0$.
    The vertical bars show the times when particle hops occur (yellow/bright lines).
    The snapshots show the configurations at the marked times (black/white indicates a particle/hole).
    }
    \label{fig: sm_traj_ssep}
\end{figure}

Figures \ref{fig: sm_traj_east}, \ref{fig: sm_traj_ssep} demonstrate trajectories sampled for the 2D East and the 2D SSEP respectively, using the approximate Doob dynamics.
The vertical plot in the left shows a light yellow line each time a transition occurs, and the right panels demonstrate a configuration snapshot at some points in time, as marked in the figures. 
We show (a) an active trajectory, (b) a trajectory close to the transition point $s_{c}$ and (c) an inactive trajectory.

\section{Comparison to other methods}
A full detailed comparison to the other methods via their numerical implementation is beyond the scope of this work.
However, we will provide a brief discussion explaining how this approach compares to other approaches.
One popular approach to estimating large deviations is cloning methods.
This approach has an exponential cost in system size, and is known to suffer from bias. 
This is most apparent around the transition point, meaning that while the method allows for a way to probe dynamical phase transitions, it is not reliable for an accurate finite size scaling analysis as performed here.
Another popular approach is trajectory sampling methods.
While this approach is unbiased, it suffers an exponential cost in both time and space.
Methods such as transition path sampling \cite{Bolhuis2002} can be used to hinder the cost, reducing the exponent in the exponential. 
Nevertheless, the cost is still exponential, and can be problematic where large times are required.

Tensor network approaches estimate the long time statistics of the dynamics by directly targetting the maximal eigenvalue of a deformed Markov generator.
Each iteration of the optimization methods scales only polynomially in system size for a fixed bond dimension.
In the case of time evolution, the number of iterations required to reach convergence is expected to scale as the inverse gap between the two leading eigenvalues of the deformed Markov generator.
Furthermore, whereas the required bond dimension for a fixed precision is not known a priori, PEPS allow for a controlled way to systematically increase the accuracy of the method by increasing 
the bond dimension at a cost which is only polynomial in bond dimension.
As shown in these works, the PEPS can be combined % implemented 
with trajectory sampling algorithms.
This allows us to approximate the most optimal sampling dynamics at a cost which again scales only polynomially in both space and PEPS bond dimension.
In practice, the dynamics is only approximate, and errors will still exponentially propagate in time.
However, the prefactor is hugely reduced from the usual setting.

\end{appendix}
\end{document}